\documentclass[prc,twocolumn,showpacs,floatfix,aps,10pt,nofootinbib]{revtex4-1}  %

\usepackage{graphicx}
\usepackage{bm}

\begin{document}
\title{Study of three-neutron bound and continuum states}
\author{Souichi Ishikawa}  \email[E-mail:]{ishikawa@hosei.ac.jp}
\affiliation{
Science Research Center, Hosei University, 
2-17-1 Fujimi, Chiyoda, Tokyo 102-8160, Japan
} 

\date{\today}

\begin{abstract} 
The three-neutron ($3n$) system is studied by numerical calculations with the Faddeev three-body formalism for a realistic nucleon-nucleon (NN) potential.   
A response function for the transition from ${}^3\mathrm{H}$ to $3n$ continuum states by an isospin excitation operator is calculated, from which no evidence of $3n$ resonance state is found.  
Different methods to extrapolate the $3n$ energy from bound state energies with an extra attractive effect to the NN potential are examined.  
While extrapolations with attractive effects by enhanced NN potentials or three-body potentials result the non-existence of $3n$ resonance states, one by external trapping potentials leads to a positive $3n$ energy, which may be considered as a resonance state. 
It is found that  this contradiction is due to a general defect of the trapping method.
\end{abstract}


\maketitle

\section{Introduction}

Studies of few-neutron systems are expected to enrich our knowledge of the interaction among neutrons, which is essential to microscopic understanding of  neuron-rich nuclei and neutron matter. 
For neutron-neutron ($nn$) system, there exists a virtual state, which results a peak in energy spectra of reactions leaving two neutrons in the final state, such as $\pi^- d \rightarrow nn\gamma$ reaction, and gives us information of the ${}^{1}\mathrm{S}_{0}$ scattering length (see e.g., Ref. \cite{Ho98}). 
There has been no conclusive evidence for the existing of bound or resonance state in $3n$ (e.g., Ref. \cite{Mi80}) and four-neutron ($4n$) systems, besides a few experimental suggestions for a resonant $4n$ state \cite{Ma02,Ki16}. 

Recently, the existence of $3n$ as well as $4n$ resonance states was indicated by quantum Monte Carlo calculations \cite{Ga17} and no-core Gamow shell model calculations \cite{Li19}. 
However, these calculations contradict with previous calculations \cite{He02,La05,De18a,De18b} (see also Refs. \cite{De19,Ga19,Tr17}) by the Faddeev-type method \cite{Fa61},  which showed that complex energy eigenvalues of the Hamiltonians of the systems are too far from the real energy axis to give any effect as a resonance. 
The aim of this paper is to clarify a reason of this discrepancy by performing Faddeev $3n$ calculations for continuum and bound states.

In Refs. \cite{He02,La05}, the Faddeev equations \cite{Fa61} were solved  in combination with the complex scaling method, from which one can obtain a complex energy eigenvalue of the system, $E_r - i \frac{\Gamma}{2}$ with $E_r$ being the resonance energy and $\Gamma$ the width. 
On the other hand, in Ref. \cite{De18b}, the transition amplitude for $3n \to 3n$ scattering was calculated at positive real energies, from which the pole position of the amplitude in the complex energy plane is evaluated.

Since the $3n \to 3n$ scattering is not able to perform as a laboratory experiment, in the present paper, I will study one of possible realization of $3n$ continuum systems, namely the charge exchange reaction:  ${}^3\mathrm{H}(n,p)3n$.  
Actually, there have been some experimental works for its mirror reaction,  ${}^3\mathrm{He}(p,n)3p$ \cite{Pa98,Wa08}. 
In Ref. \cite{Is18}, this reaction was studied in a plane wave impulse approximation (PWIA), in which response functions of the transition from ${}^3\mathrm{He}$ to the three-proton ($3p$) state by spin-isospin transition operators were calculated in  the Faddeev three-body formalism.

The response function is written as the imaginary part of a matrix element of the Green's function for the $3n$ Hamiltonian (see Eq. (\ref{eq:R_Greens-function}) below). 
A complex eigenvalue of the Hamiltonian corresponds to the pole of the Green's function in the complex energy plane.
If the pole is close to positive energy axis in the fourth quadrant so that  the response function has a peak as a function of the (real) energy,  we may recognize that a resonance state exists.

In the above mentioned calculations, to realize a bound state or a resonance state artificially, an additional attractive effect is given on the original Hamiltonian either by enhancing the $nn$ interaction  \cite{He02,La05,De18b}, by introducing a three-body force  \cite{La05}, or by introducing an external potential that confines the neutrons in a trap \cite{Ga17,Li19}.  
Energies are calculated with modifying the strength of the attractive effect, from which the energy for the original Hamiltonian is extracted. 

In Sec. \ref{sec:Response-function}, three-body calculations of the response function will be described.
In Sec. \ref{sec:calculation}, results of the response function as well as the $3n$ binding energy for modified $3n$ Hamiltonians  are presented, and the extrapolation methods will be examined. 
In Sec. \ref{sec:2n-Gaussian}, results of $3n$ calculations will be interpreted in a simple two-body system.
Summary will be gin in Sec. \ref{sec:summary}.

\section{Response function}
\label{sec:Response-function}
%

In this paper, I will study a response function corresponding to ${}^{3}$H bound state to $3n$ continuum state by an isospin excitation operator:
\begin{equation}
\hat{O}(Q) = \sum_{i=1}^{3} e^{i Q \hat{\boldsymbol{z}}\cdot \boldsymbol{r}_i}  t^{(-)}_{i},
\label{eq:O_c}
\end{equation}
where $Q$ is the momentum transfer, $t^{(-)}_{i}$ an isospin operator that transforms the proton $i$ in ${}^3$H to neutron $i$ in the final $3n$ state, $\boldsymbol{r}_i$  the coordinate vector in the three-nucleon center of mass (c.m.) system of the particle $i$. 
This is one of three transition operators to used in PWIA analysis of the ${}^3\mathrm{H}(n,p)3n$ reaction.

First, I introduce the $3n$ Hamiltonian in the c.m. system,
\begin{equation}
H_{3n} = H_0 + \sum_{i} V^{\mathrm{(2B)}}_{i} + \sum_{i} V^{\mathrm{(3B)}}_{i},
\end{equation}
where $H_0$ is the three-body kinetic energy operator, $V^{\mathrm{(2B)}}_{i}$ is an two-body potential between particles $j$ and $k$, and $V^{\mathrm{(3B)}}_{i}$ is a three-body potential (3BP) that is symmetric with respect to particles $j$ and $k$.

Let $\vert \Psi^{(\pm)}(q,p;J^\pi)\rangle$ be an eigenstate of the $3n$ Hamiltonian ${H} _{3n}(J^\pi)$ for the total angular momentum $J$ and parity $\pi$ state associated with an asymptotic $3n$-state, in which the relative momentum between two neutrons is $q$, the momentum of the third neutron with respect to c.m. of the neutron-pair  is $p$.
The superscript $(\pm)$ expresses the outgoing $(+)$ or incoming $(-)$  boundary condition.

The eigenvalue problems is written as 
\begin{equation}
{H}_{3n}(J^\pi) \vert  \Psi^{(\pm)} \left( q,p;J^\pi\right) \rangle
= E(q,p)  \vert  \Psi^{(\pm)} \left( q,p;J^\pi\right) \rangle,
\end{equation}
with 
\begin{equation}
E(q,p)  = \frac{q^2}{m}  + \frac{3 p^2}{4 m},
\end{equation}
where $m$ is the mass of the neutron.

A response function corresponding to the transition from the ${}^3$H bound state, $\vert \Psi_t \rangle$, to $3n$-continuum states with energy $E$ by an operator $\hat{O}(Q)$ is written as
\begin{eqnarray}
R(E,Q;J^\pi) &=&  \int dq dp
\left\vert T(q,p;Q,J^\pi) \right\vert^2 
\delta\left(E-E(q,p)\right),
\cr &&
\label{eq:R_E}
\end{eqnarray}
where  the transition amplitude is defined by 
\begin{equation}
T(q,p;Q,J^\pi) = 
\langle  \Psi^{(-)} \left( q,p;J^\pi\right) \vert \hat{O}(Q) \vert 
\Psi_t  \rangle.
\label{eq:T_amp}
\end{equation}
Using the completeness of the $3n$ states, we have
\begin{eqnarray}
&& R(E,Q;J^\pi)
\cr
&=&
\langle \Psi_t \vert\hat{O}^{\dagger}(Q) \delta(E-{H}_{3n}(J^\pi)) \hat{O}(Q)  \vert \Psi_t \rangle
\cr
&=&
- \frac{1}{\pi}  \mathrm{Im}  
\langle \Psi_t \vert\hat{O}^{\dagger} (Q)
 \frac{1}{E+ i \epsilon -{H}_{3n}(J^\pi)} \hat{O}(Q)  \vert \Psi_t \rangle.
\label{eq:R_Greens-function}
\end{eqnarray}

Here, I introduce a wave function $\vert \Xi (Q,J^\pi)\rangle$ describing the disintegration process,   
\begin{equation}
\vert \Xi(Q,J^\pi) \rangle = \frac{1}{E+i \epsilon - {H}_{3n}(J^\pi)} \hat{O}(Q) \vert \Psi_t \rangle.
\label{eq:Psi_def}
\end{equation}
Adapting the Faddeev theory to solve Eq. (\ref{eq:Psi_def}), a three-body wave function $\vert \Xi \rangle$ is decomposed into three (Faddeev) components:
\begin{equation}
\vert \Xi \rangle= \vert\Phi^{(1)}\rangle +\vert \Phi^{(2)}\rangle 
  + \vert \Phi^{(3)}\rangle,  
\label{eq:Fad-dec}
\end{equation}
where I drop the arguments $Q$ and $J^\pi$ for simplicity.
Corresponding to this decomposition, the operator $\hat{O}$ is decomposed into three components:
\begin{equation}
\hat{O} = \hat{O}_{1} + \hat{O}_{2} + \hat{O}_{3},     
\label{eq:O-dec}
\end{equation}
with the condition that  $\hat{O}_{i}$ is symmetric with respect to the exchange of $j$ and $k$.
Then Faddeev equations  read:
\begin{eqnarray}
\vert\Phi^{(1)}\rangle &=& {G}_1(E) \hat{O}_{1} \vert \Psi_t \rangle 
 +{G}_1(E) V^{\mathrm{(2B)}}_1 \vert \Phi^{(2)} + \Phi^{(3)} \rangle   
\cr
&&+{G}_1(E) V^{\mathrm{(3B)}}_1 
\vert \Phi^{(1)} + \Phi^{(2)} + \Phi^{(3)}  \rangle, 
\cr
&&
\text{(and ~cyclic~ permutations)},
\label{eq:Fad-eq}
\end{eqnarray}
where the operator ${G}_i(E)$ is a channel Green's function defined as
\begin{equation}
{G}_i (E) \equiv 
\frac1{ E + \imath\varepsilon - H_0 - V^{\mathrm{(2B)}}_i}.
\label{eq:channel-Green}
\end{equation}
The Faddeev equations, Eq. (\ref{eq:Fad-eq}), are solved  as integral equations in coordinate space, whose formal and technical details  are essentially same as those used for the nucleon-deuteron scattering \cite{Is03,Is09} and three alpha-particles \cite{Is13} problems.
The amplitude, Eq. (\ref{eq:T_amp}), is calculated from the solution.

In the present work, I will use the Argonne V$_{18}$ model (AV18) \cite{Wi95} for the NN potential taking partial waves with angular momenta $j \le 4$. 
The ${}^{3}$H wave function for the initial state is calculated with  AV18 and the Brazil  2$\pi$-exchange type three-nucleon potential, BR-${\cal O}(q^4)$ in Ref. \cite{Is07a}.

\section{$3n$ Calculations}
\label{sec:calculation}

I will consider the transition from the ${}^{3}$H ground state to $3n (\frac{3}{2}^{-} )$ continuum state.
This final state was reported to be the most preferable in a sense that the modification of the original nuclear interaction to produce a resonance state could be minimal \cite{He02,La05}, and is considered to be the state found in Ref. \cite{Ga17}. 

\begin{figure}[tb]
\includegraphics[width=0.65\columnwidth,angle=-90]{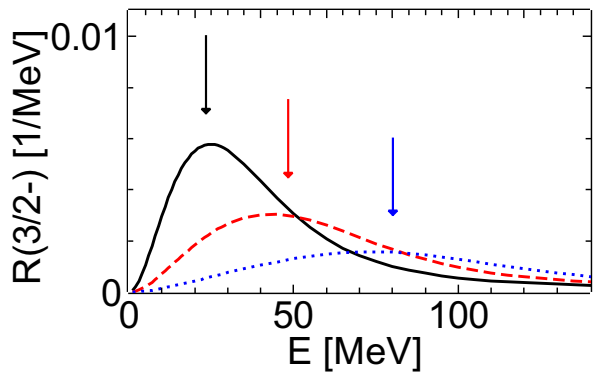}
\caption{(Color online) 
Energy dependence of the Response function $R(E,Q; \frac{3}{2}^{-})$  for $Q=300$ MeV/c (black solid curve), $Q=400$ MeV/c (red dashed curve),  and $Q=500$ MeV/c (blue dotted curve) calculated with AV18. 
The black, red, and blue arrows indicate the energies given by Eq. (\ref{eq:e-quasi-free}) for $Q=300$ MeV/c, 400 MeV/c, and 500 MeV/c, respectively.  
\label{fig1:RC-jp4-av18}
}
\end{figure}

Fig. \ref{fig1:RC-jp4-av18} displays the response functions $R(E,Q; \frac{3}{2}^{-})$ for $Q=300$, 400, and 500 MeV/c calculated with AV18 as functions of $E$.
The figure shows that the response functions have a peak at the energy, which varies with $Q$.
The vertical arrows indicate the energies calculated by 
\begin{equation}
E = \frac{Q^2}{2m} - B({}^{3}\mathrm{H}) - \frac{Q^2}{6m},
\label{eq:e-quasi-free}
\end{equation}
where $B({}^{3}\mathrm{H})$ is the ${}^{3}$H binding energy.
This value means that the momentum $Q$ is absorbed by one neutron, which leads to a quasi-free process. 
The peak energy of the response function for each of $Q$ almost coincides with the one given by Eq. (\ref{eq:e-quasi-free}), which shows that the peaks of $R(E,Q)$ are not due to a resonance pole. 
   
In the following, I will examine three extrapolation procedures with giving additional attractions to the $3n$ Hamiltonian: 
(i)  by multiplying a factor to the $nn$ potential to enhance an attractive contribution;
(ii) by introducing a 3BP; 
and (iii) by introducing additional potential that traps neutrons around their center of mass.

{(i)}
The factor multiplied to the $nn$ potential will be denoted by $(1-\alpha)$. 
Since a rather small value of $\alpha$, e.g. $-0.080$ for AV18, makes $nn({}^{1}\mathrm{S}_0)$ system bind \cite{La05}, the factor will be multiplied only to ${}^3\mathrm{P}_2$-${}^3\mathrm{F}_2$ partial wave component of the $nn$ potential, which is known to be attractive. 
In this notation, a negative value of $\alpha$ gives an attractive effect.
In fact, a $nn({}^3\mathrm{P}_2$-${}^3\mathrm{F}_2)$ bound state exists for $\alpha < -3.39$, and a $3n (\frac{3}{2}^{-})$ bound state does for $\alpha<-2.98$. 
These values agree with those obtained in Ref. \cite{La05}.

\begin{figure}[tb]
\includegraphics[width=0.32\columnwidth,angle=-90]{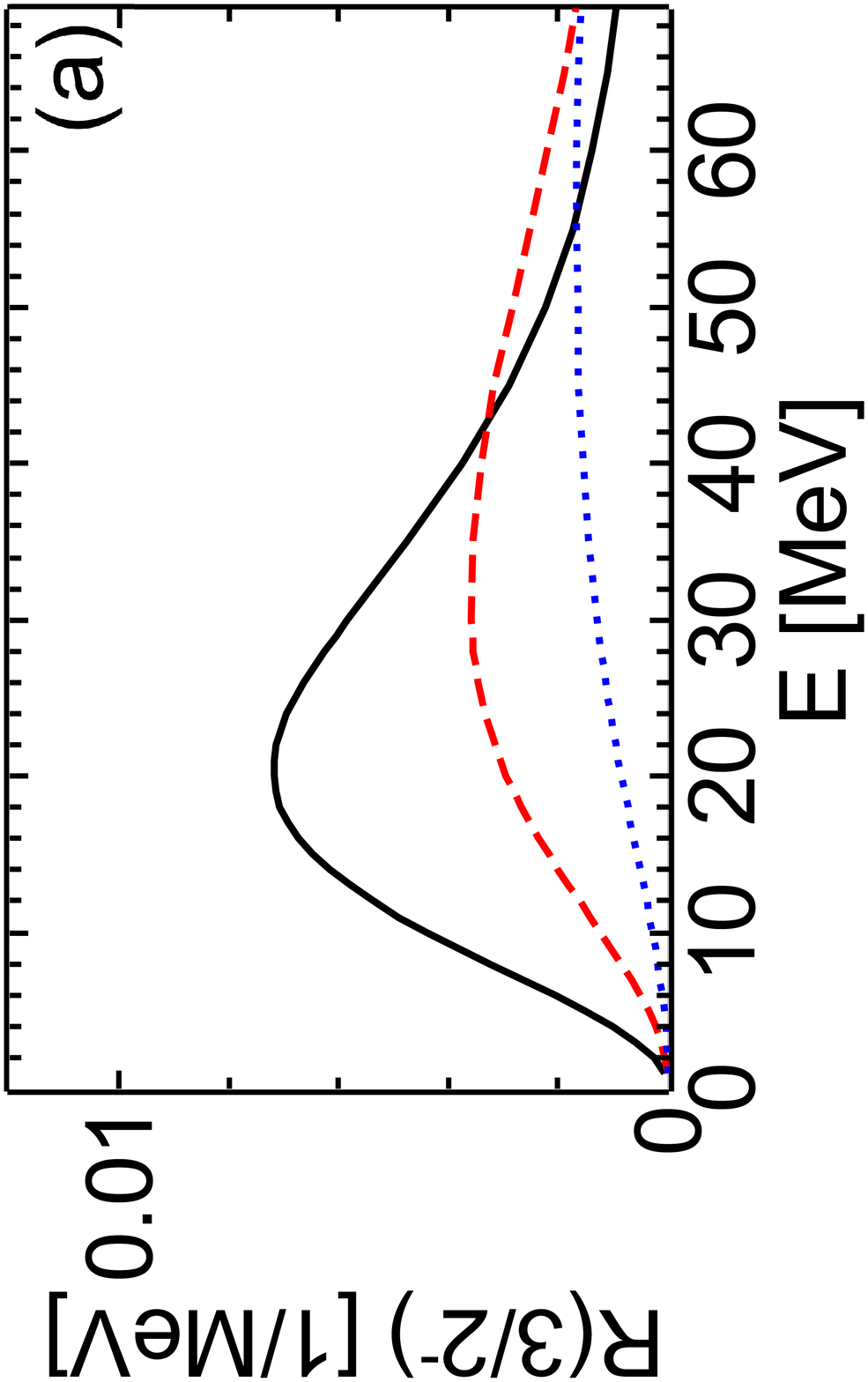}
\includegraphics[width=0.32\columnwidth,angle=-90]{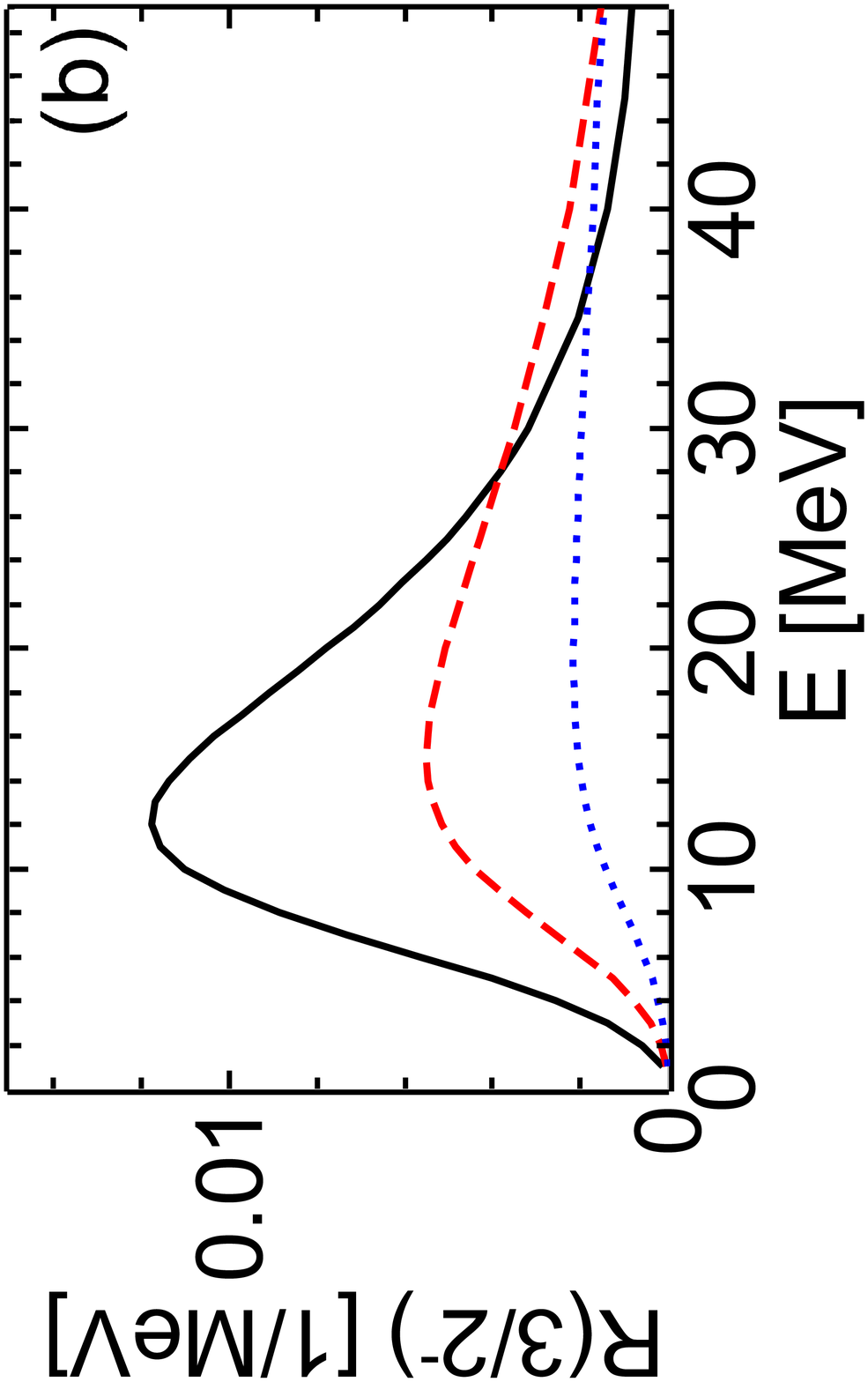}
\includegraphics[width=0.32\columnwidth,angle=-90]{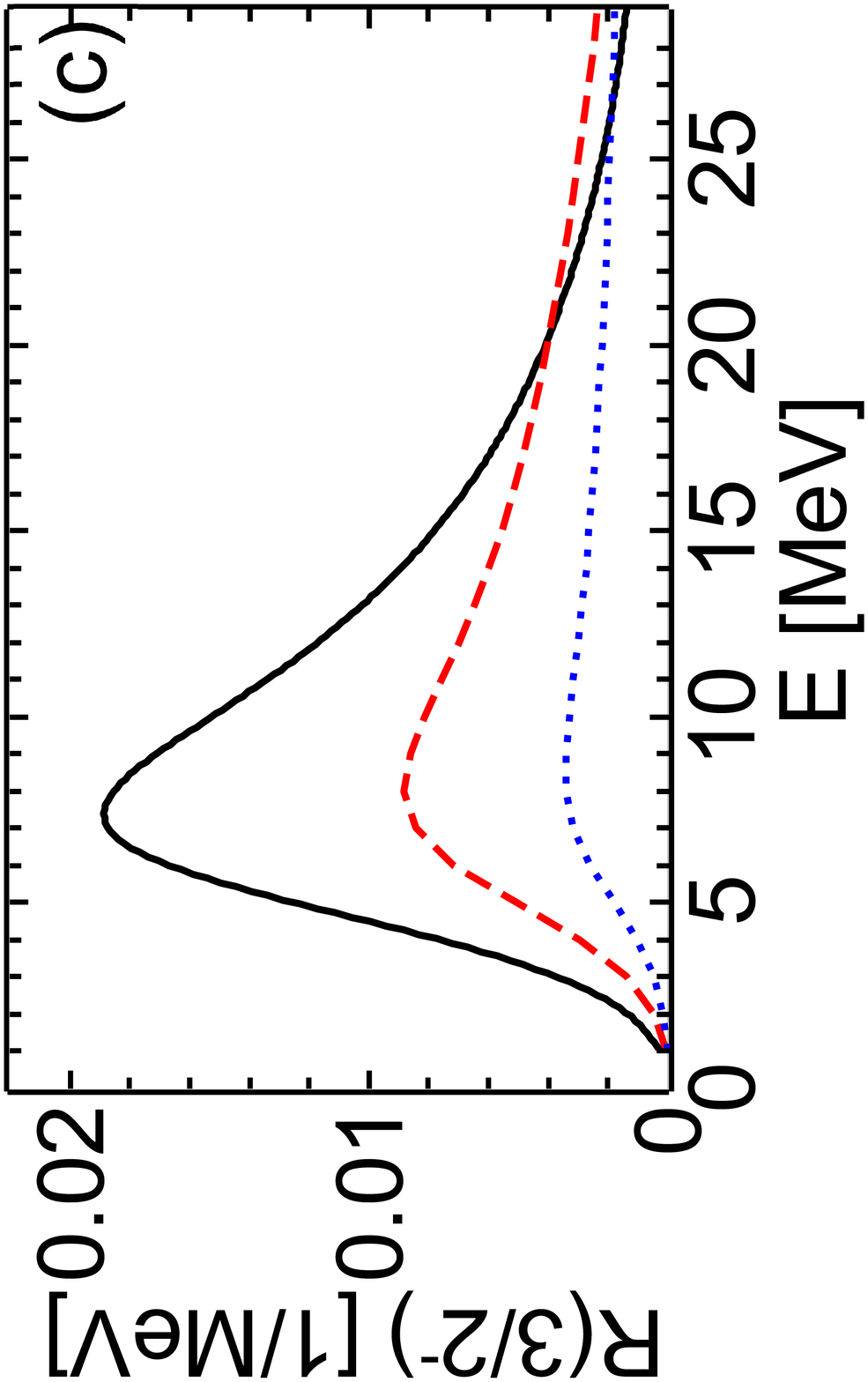}
\includegraphics[width=0.32\columnwidth,angle=-90]{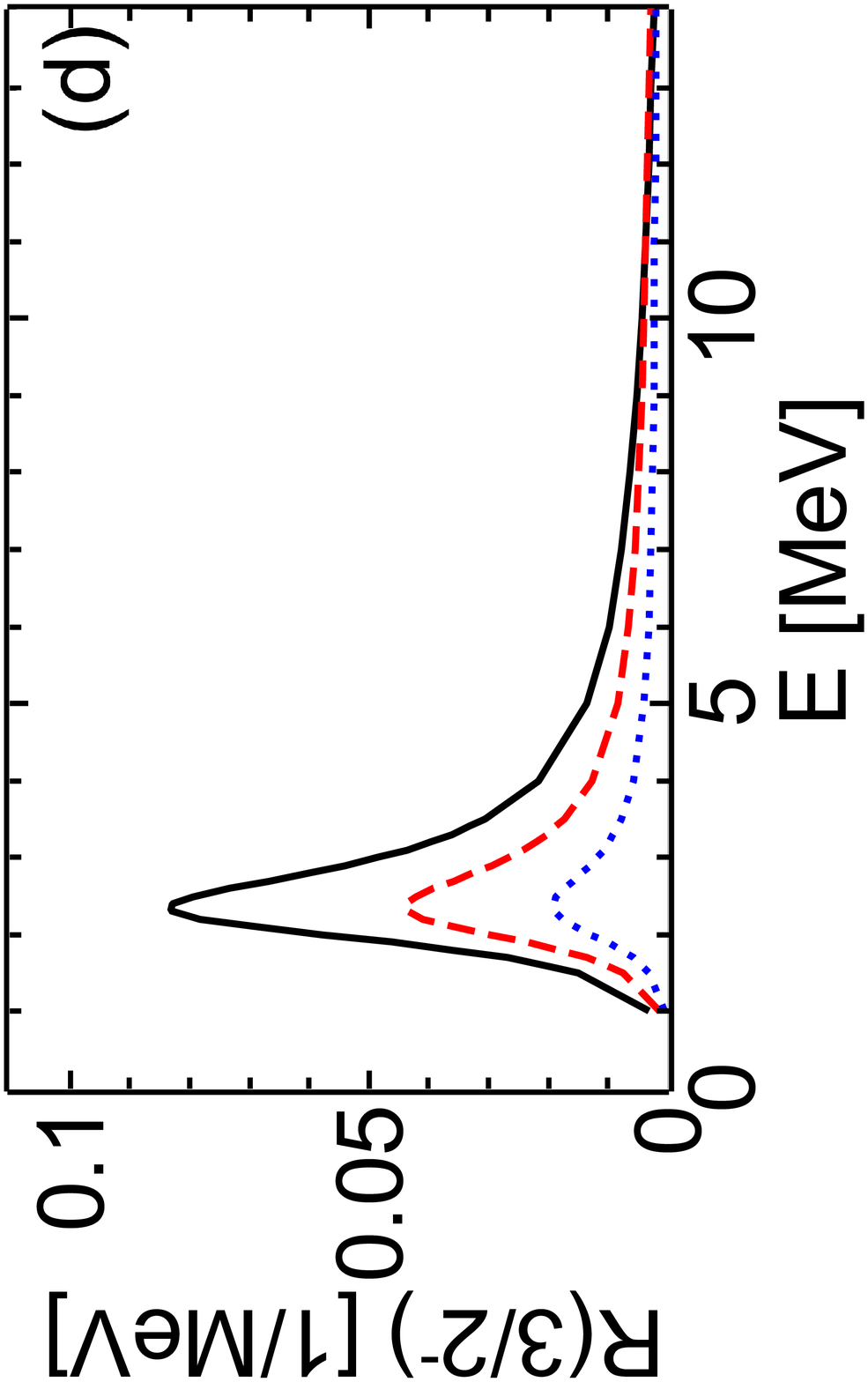}
\caption{(Color online) 
Energy dependence of the response function $R(E,Q; \frac{3}{2}^{-})$ calculated by ${}^3\mathrm{P}_2$-${}^3\mathrm{F}_2$ modified AV18 with (a) $\alpha = -1.0$, (b) $\alpha =-2.0$,  (c) $\alpha =-2.4$, and (d) $\alpha =-2.8$.
In each figure, black solid, red dashed, and blue dotted curves denote for  $Q=300$ MeV/c, 400 MeV/c, and 500 MeV/c, respectively.
\label{fig2:RC-jp4-3pf2}
}
\end{figure}

\begin{figure}[tb]
\includegraphics[width=0.65\columnwidth,angle=-90]{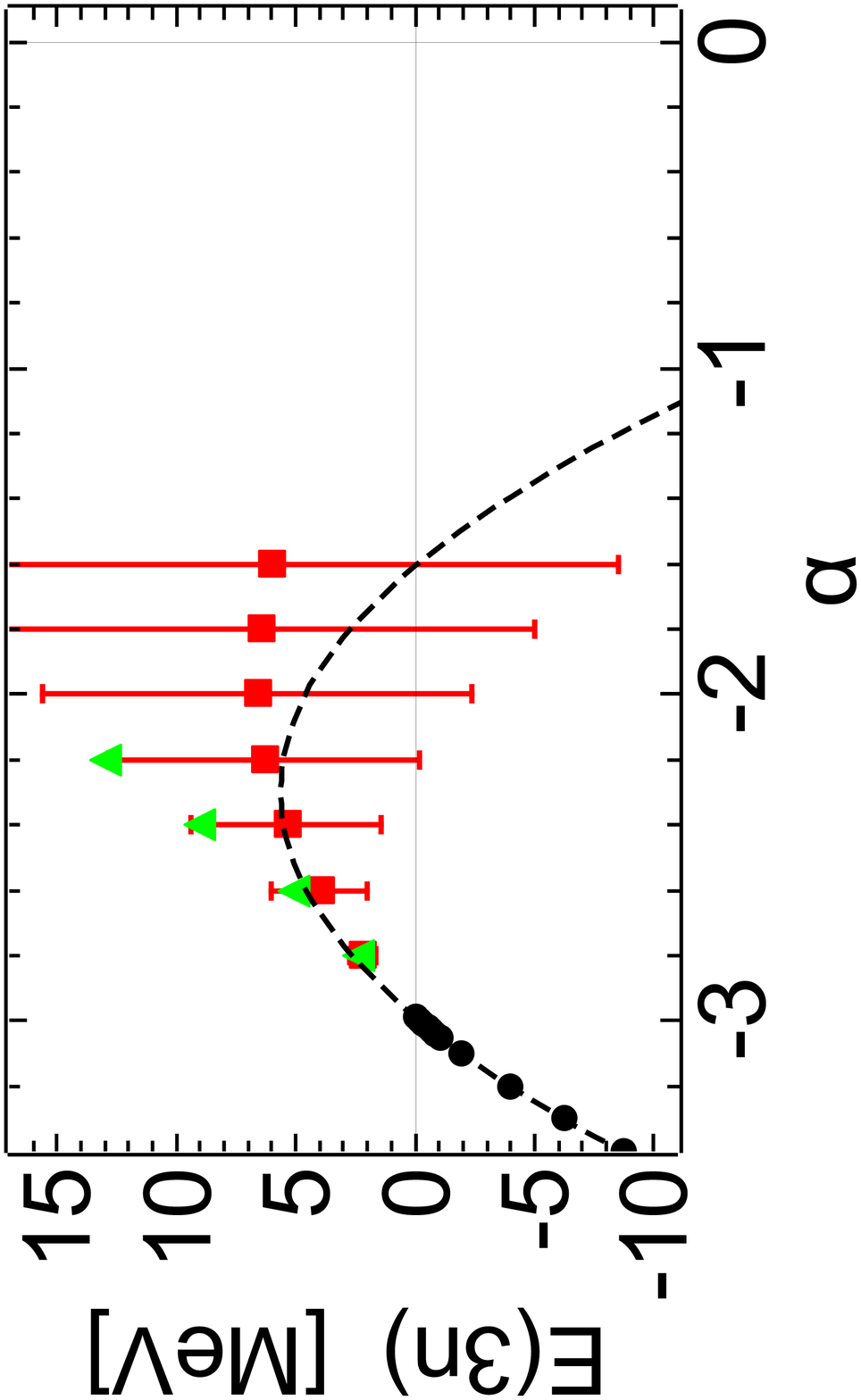}
\caption{(Color online) 
Calculated $3n$ energies for AV18 as functions of the factor $\alpha$. 
Black circles are calculated values of the $3n$ binding energy.
Black dashed curves are obtained by fitting the $3n$ binding energy. 
Green triangles are the peak energies of the response functions. 
Red squares extracted values of $E_r$ with error bars being $\pm \frac12 \Gamma$ from the response functions using Eq. (\ref{eq:Lorentz-fit}).
\label{fig3:3n-be-3pf2}
}
\end{figure}

The response functions for $Q=300$, 400, and 500 MeV/c calculated by ${}^3\mathrm{P}_2$-${}^3\mathrm{F}_2$ modified AV18 with $\alpha=-1.0$, $-2.0$, $-2.4$, and $-2.8$ are displayed in Fig. \ref{fig2:RC-jp4-3pf2}. 
As the attractive effect becomes larger,  the $Q$-dependence of the peak energy does weaker. 
For  $\alpha < -2.0$, the peak energies are almost $Q$-independent, and are plotted as green triangles  in Fig. \ref{fig3:3n-be-3pf2} as a function of $\alpha$.

Since the peak energy may not necessary be the resonance energy, I will test to fit the response function by the following expression: 
\begin{eqnarray}
R(E) &=& \frac{b(E-E_r)+ c \Gamma}{(E-E_r)^2 + \Gamma^2/4}
\cr
&& + a_0 + a_1 (E-E_r) + a_2 (E-E_r)^2,
\label{eq:Lorentz-fit}
\end{eqnarray}
which has a form of  the Lorentz function taking into account some asymmetric effects.  
The parameters $E_r$, $\Gamma$, $a_n (n=0,1,2)$, $b$, and $c$ are obtained from the response function for $E \le 30$ MeV. 
Here the complex value $E_r - i \frac{\Gamma}{2}$ could be a pole energy of the Green's function in the complex plane. 
Extracted values of $E_r$ and $\Gamma$ are $Q$-independent for $-2.7 \le \alpha \le -1.6$. 
In Fig. \ref{fig3:3n-be-3pf2}, extracted values of $E_r$ are plotted as red squares with error bars being $\pm \frac{1}{2} \Gamma$. 
As the attractive effect is reduced, $E_r$ increases and stays at about 6.5 MeV with increasing the width, which reaches about $\Gamma=30$ MeV at $\alpha=-1.6$. 
Numerical errors of $E_r$ and $\Gamma$ in the extraction are small  as  $10^{-3}$ MeV for $\alpha=-2.7$, and increase as the magnitude of $\alpha$ becomes small: about 0.4 MeV for $\alpha=-1.6$. 
Extracted values of $E_r$ and $\Gamma$ are $Q$-dependent for $\alpha > -1.6$, which indicates that the complex value $E_r - i \frac{\Gamma}{2}$ is away from the real energy axis.
These tendencies in the obtained values of $(E_r, {\Gamma})$ are similar to those from the $3n \to 3n$ amplitude with ${}^3\mathrm{P}_2$-${}^3\mathrm{F}_2$ modified $nn$ potentials in Ref. \cite{De18b}.

Calculated values of  the $3n (\frac{3}{2}^{-})$ binding energy for $\alpha < -3.0$ are plotted as black circles in Fig. \ref{fig3:3n-be-3pf2}. 
As shown in the figure, both of the peak energies and the extracted $E_r$ look to be smoothly connected to the binding energies. 
An extrapolation is performed by fitting the $3n$ binding energies with a quadratic polynomial of $\alpha$, which is  chosen just for its simplicity.
The result is plotted as black dashed curve in the figure, which approximately follows the extracted $E_r$ rather than the peak energy.


{(ii)}
The introduction of an attractive 3BP is another way to bring an extra attractive effect.
In Ref. \cite{Is18}, the response functions for spin-isospin transitions from ${}^{3}$He ground state to $3p$ continuum state  are calculated with introducing a 3BP to produce a $3p$ resonance mandatory. 
Here, I will apply the same functional form of  3BP, which was taken from  Ref. \cite{Hi16}: 
\begin{equation}
V^{\mathrm{(3B)}}_i = \frac{1}{3}\sum_{n=1}^{2} W_n e^{-(r_{ij}^2+r_{jk}^2+r_{ki}^2)/b_n^2}, 
\label{eq:g-3bp}
\end{equation}
where  $r_{ij}$ is the distance between the $i$-th and $j$-th neutrons. 
Note that this form of $V^{\mathrm{(3B)}}_i$ is totally symmetric under the particle exchange and the total 3BP is obtained by summing up all cyclic permutations of Eq. (\ref{eq:g-3bp}), and the factor $1/3$ arises because of that. 
The range parameters and the strength parameters of the shorter range term are the same ones as used in Refs. \cite{Hi04,Hi16}:  $b_1=4.0$ fm, $b_2=0.75$ fm, and $W_2=35.0$ MeV.

When the 3BP is applied to $3n (\frac{3}{2}^{-})$ state with AV18, there is at least one $3n$ bound state for the attractive strength $W_1 < -80$ MeV. 
The value $W_1 = -80$ MeV contrasts with $W_1=-2.55$ MeV that is determined to reproduce  ${}^{3}$H binding energy \cite{Is18}.
Also, it is noted that the required value of the strength parameter $W_1$ for  $4n (0^+)$ state to bind as the lower bound of the experimental value \cite{Ki16} is $-36.14$ MeV \cite{Hi16}.

\begin{figure}[tb]
\includegraphics[width=0.32\columnwidth,angle=-90]{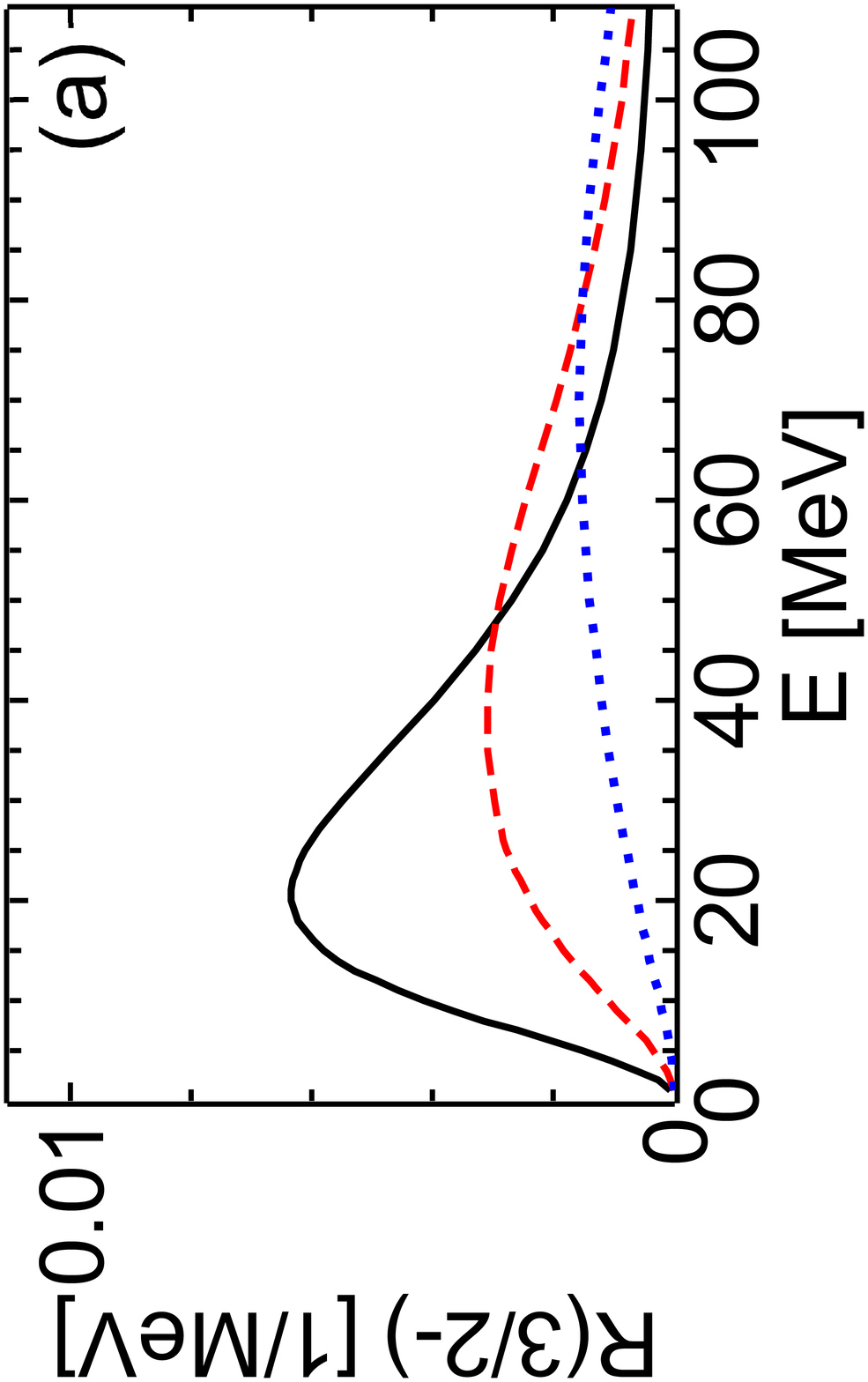}
\includegraphics[width=0.32\columnwidth,angle=-90]{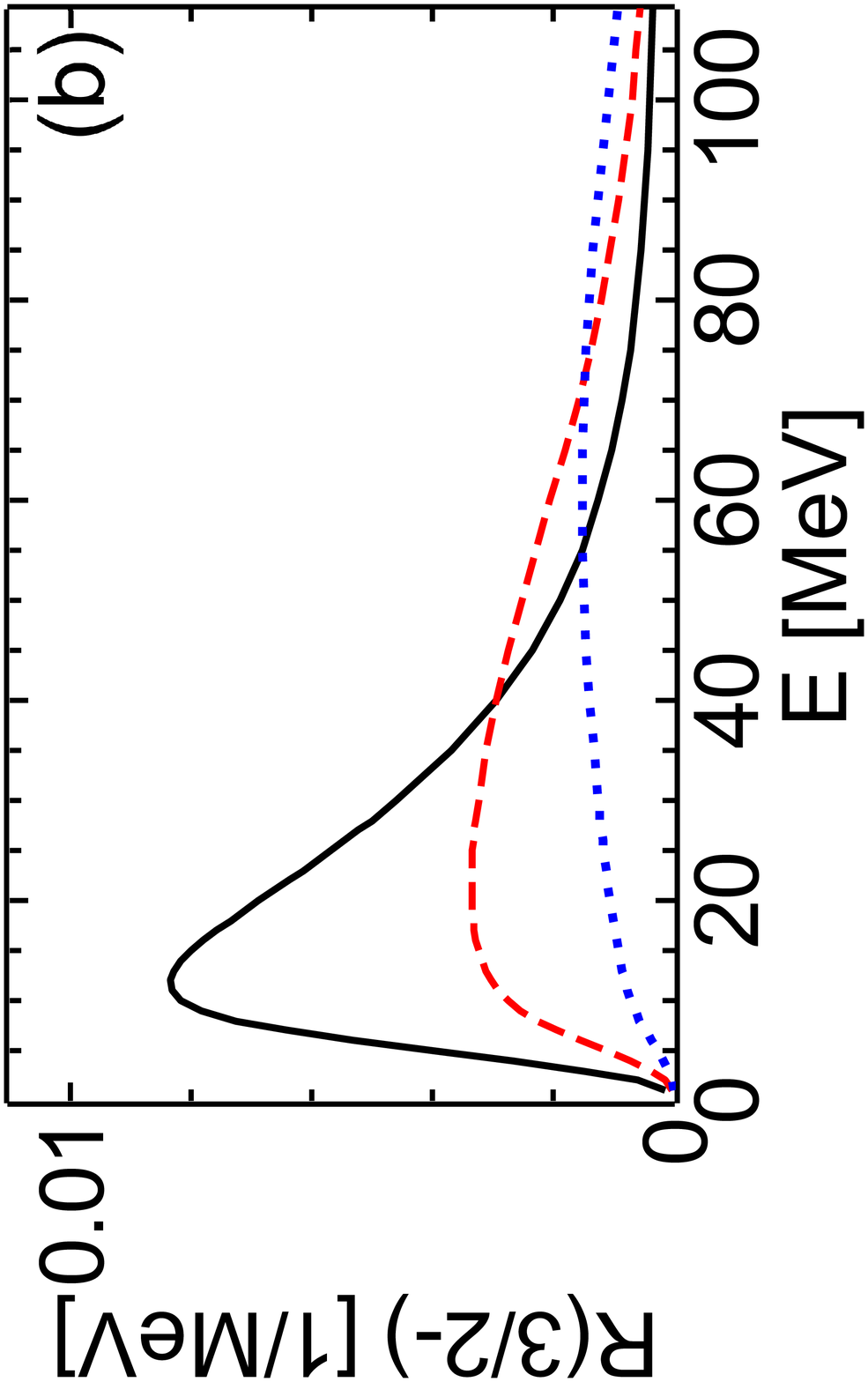}
\includegraphics[width=0.32\columnwidth,angle=-90]{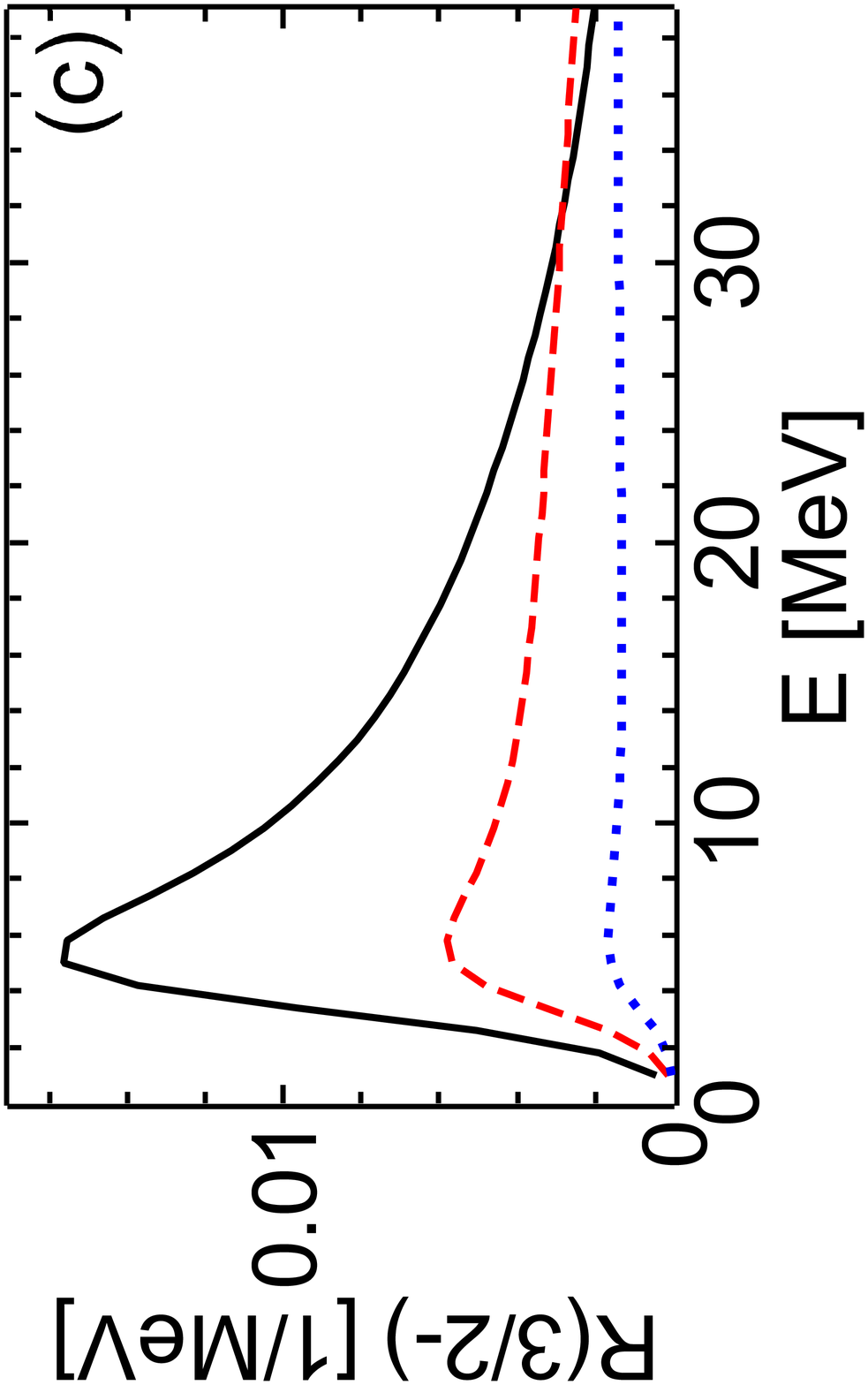}
\includegraphics[width=0.32\columnwidth,angle=-90]{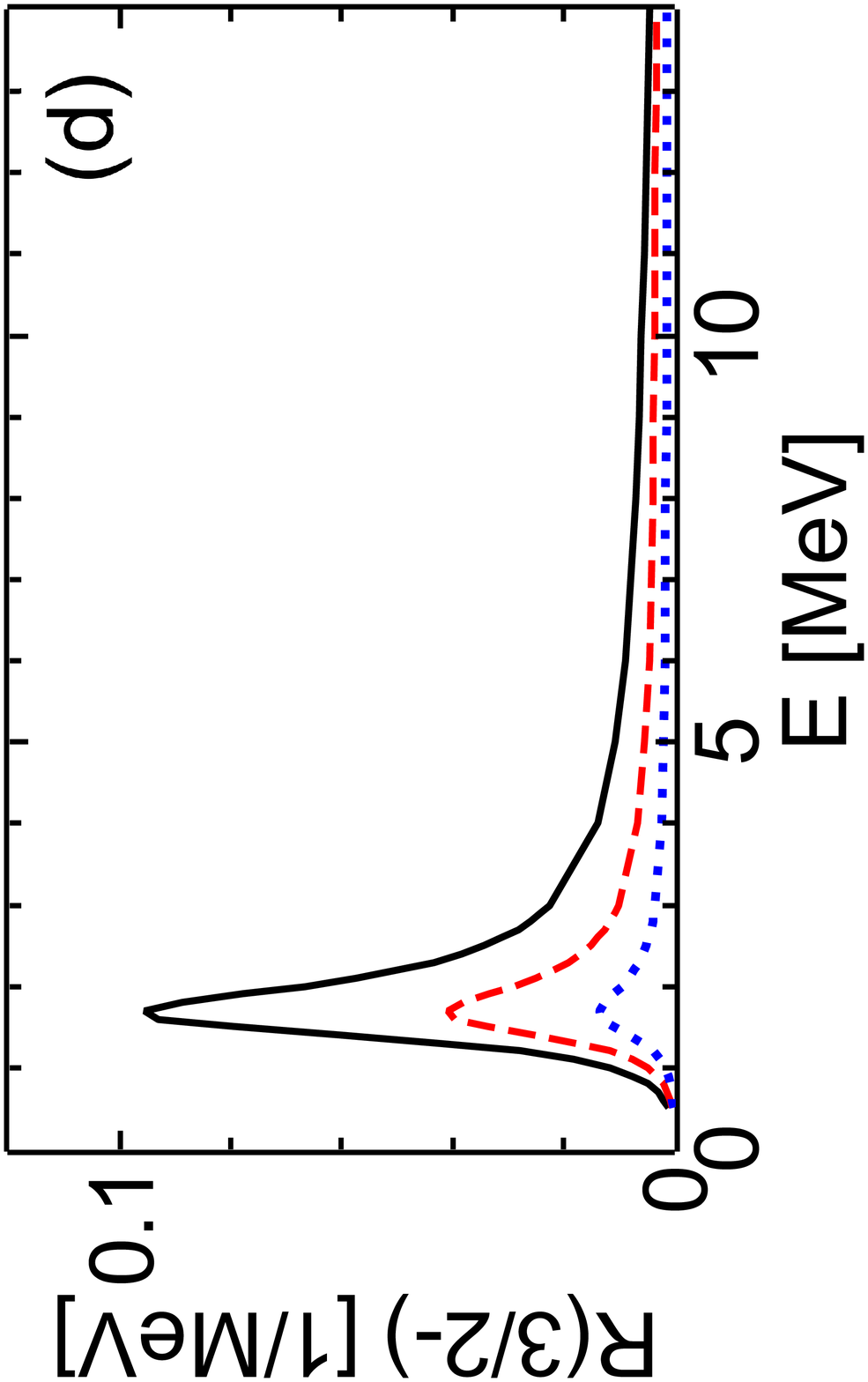}
\caption{(Color online) 
Energy dependence of the Response function $R(E,Q; \frac{3}{2}^{-})$ calculated with AV18+3BP for (a) $W_1=-10$ MeV, (b) $W_1=-30$ MeV, (c) $W_1=-50$ MeV, and (d) $W_1=-70$ (MeV).
In each figure, black solid, red dashed, and blue dotted curves denote $R(E,Q; \frac{3}{2}^{-})$ for  $Q=300$ MeV/c, 400 MeV/c, and 500 MeV/c, respectively.
\label{fig4:RC-jp4-3bp}
}
\end{figure}

\begin{figure}[tb]
\includegraphics[width=0.65\columnwidth,angle=-90]{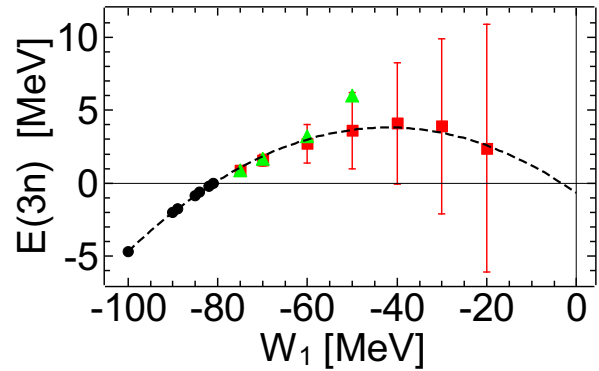}
\caption{(Color online)  
Calculated $3n$ energies as functions of the 3BP strength parameter $W_1$. 
Black circles are calculated values of the $3n$ binding energy, and black dashed curve is obtained by fitting the $3n$ binding. 
Green triangles are the peak energies of the response functions, and red squares are extracted values of $E_r$ with error bars being $\pm \frac12 \Gamma$ from the response functions using Eq. (\ref{eq:Lorentz-fit}).
\label{fig5:3n-be-3bp}
}
\end{figure}

Fig. \ref{fig4:RC-jp4-3bp} shows the response functions for $Q=300$, 400, and 500 MeV/c, calculated with AV18 plus 3BP of the strength parameter $W_1=-10$, $-30$, $-50$, and $-70$ MeV. 
The dependence of the peak energy on $Q$ becomes weak as the magnitude of $W_1$ increases. 
For  $-80~\mathrm{MeV}<W_1 < -50~\mathrm{MeV}$, the peak energies are almost $Q$-independent, and are plotted as green triangles  in Fig. \ref{fig5:3n-be-3bp}.

Extracted values of $E_r$ and $\Gamma$ from the response functions using Eq. (\ref{eq:Lorentz-fit}) are plotted in the form of  $E_r \pm \frac{1}{2} \Gamma$ for  $-80~\mathrm{MeV} < W_1 < -20$ MeV, where the values are $Q$-independent.
As the attractive effect is reduced, $E_r$ once increases and then decreases having the maximum value of about 4 MeV and  the width of about $\Gamma=8$ MeV at $W_1 = -40$ MeV. 
Numerical errors of $E_r$ and $\Gamma$ in the extraction are about $10^{-4}$ MeV for $W_1=-75$ MeV, and increase as the magnitude of the 3BP decreases:  0.1 MeV for $W_1=-20$ MeV.

The calculated values of the $3n$ binding energy are plotted as black circles in Fig. \ref{fig5:3n-be-3bp}, and a quadratic fit to these energies is shown as black dashed curve.
It is interesting to see that the fitted curve, in spite of its simple quadratic form, almost follows the extracted vales of $E_r$ using Eq. (\ref{eq:Lorentz-fit}).

The variations of $(E_r, \Gamma)$ with respect to the parameters to modify the attractive effect in the cases (i) and (ii) are very similar to the pole trajectory obtained in the previous Faddeev calculations \cite{La05, De18b}. 
In both cases, the quadratic fitting of the $3n$ binding energy  leads to a conclusion that there is no pole in the complex energy plane close to the real axis.

{(iii)}
Next, I will examine the extrapolation using a trapping potential. As in Refs. \cite{Ga17,Li19}, I use a potential of Woods-Saxon form with a radius $R_\mathrm{WS}$ and a diffuseness parameter $a_\mathrm{WS}= 0.65$ fm, 
\begin{equation}
W(r_i) = W_\mathrm{WS} \frac{1}{1+ e^{(r_i-R_\mathrm{WS})/a_\mathrm{WS}}},
\label{eq:WS-pot}
\end{equation}
where $r_i$ is the distance of $i$-th neutron from the c.m. of the $3n$ system.

For the $3n$ bound state problem with the one-body potential $W(r_i)$, Faddeev calculations are performed in a way that the  potential  $W(r_1)$ is treated as same as three-body potential, $V^{\mathrm{(3B)}}_1$ in Eq. (\ref{eq:Fad-eq}).
This treatment works well thanks to the limited range of wave functions for the bound state problem. 

\begin{figure}[tb]
\includegraphics[width=0.65\columnwidth,angle=-90]{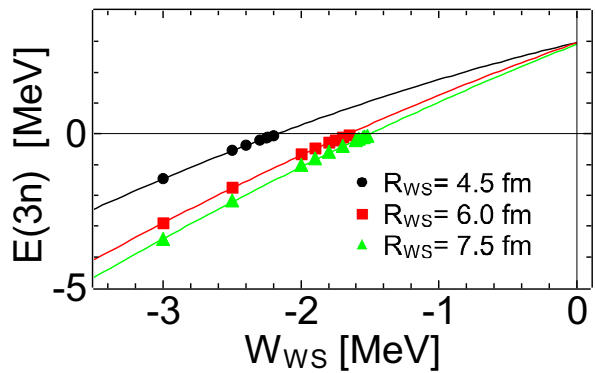}
\caption{(Color online) 
Calculated $3n$ binding energies for AV18 as functions of the strength of the trapping potential $W_\mathrm{WS}$. 
Black circles, red squares, and green triangles are calculated for the range parameter $R_\mathrm{WS}= 4.5$, 6, and 7.5 fm. 
Curves are obtained by fitting the points with a quadratic polynomial of $W_\mathrm{WS}$.
\label{fig6:3n-be-ws}
}
\end{figure}

Calculated $3n$ binding energies for some values of $R_\mathrm{WS}$ are plotted in Fig. \ref{fig6:3n-be-ws} as functions of the potential strength parameter $W_\mathrm{WS}$.
The curves are obtained by fitting the calculated energies with a quadratic polynomial of $W_\mathrm{WS}$, and are extrapolated to $W_{\mathrm{WS}}=0$ MeV.  
Extrapolated $3n$ energies with different $R_\mathrm{WS}$ values almost coincide with about 3 MeV. 
This may be the same result with one of Refs. \cite{Ga17,Li19}, which suggests the existing of $3n$ resonance.

On the other hand, the results of the extrapolation methods (i) and (ii) demonstrate that the extrapolated $3n$ complex energy for the AV18 potential has a large negative imaginary part, and the real part of the energy may be negative, which indicates the non-existing of $3n$ resonance. 

These contradictive results throw some doubt on the reliability of the extrapolation method by a trapping potential.
In the next section, a possible reason for this will be discussed.

\section{$2n$ systems with Gaussian potential}
\label{sec:2n-Gaussian}

Having in mind that a naive picture of $3n ({\frac{3}{2}}^{-})$ state is a two-body system of the spin-singlet $nn$ pair (dineutron) and the neutron in P-wave ($L=1$) state, I will apply the extrapolation method (iii) in the previous section to a two-body (two-neutron) P-wave system. 
In general, a P-wave resonance state may occur because of an attractive potential pocket within an exterior barrier caused by the centrifugal potential. 
Here, I define an effective potential $V_\mathrm{eff}(x;W_{\mathrm{WS}})$ as the sum of an attractive Gaussian potential, the P-wave  centrifugal potential, and the trapping potential, Eq. (\ref{eq:WS-pot}): 
\begin{eqnarray}
V_\mathrm{eff} (x; W_{\mathrm{WS}}) &=& 
 v_\mathrm{G} \exp\left( -(x/r_\mathrm{G})^2  \right)
+ \frac{\hbar^2 L(L+1)}{m x^2}
\cr
&& 
+ \sum_{i=1,2} W(r_i), 
\label{eq:V-eff}
\end{eqnarray}
where $x$ is the distance between two particles.
In this study, I take the parameters of the Gaussian potentials as $r_\mathrm{G}=2.5$ fm and $v_\mathrm{G}=-50 ~\mathrm{MeV}$.
Calculated P-wave scattering phase shift for this Gaussian potential takes a maximum of about  $70^{\circ}$ starting from $0^{\circ}$ at zero energy, which means that the system does not have a resonance state. 

\begin{figure}[tb]
\includegraphics[width=0.34\columnwidth,angle=-90]{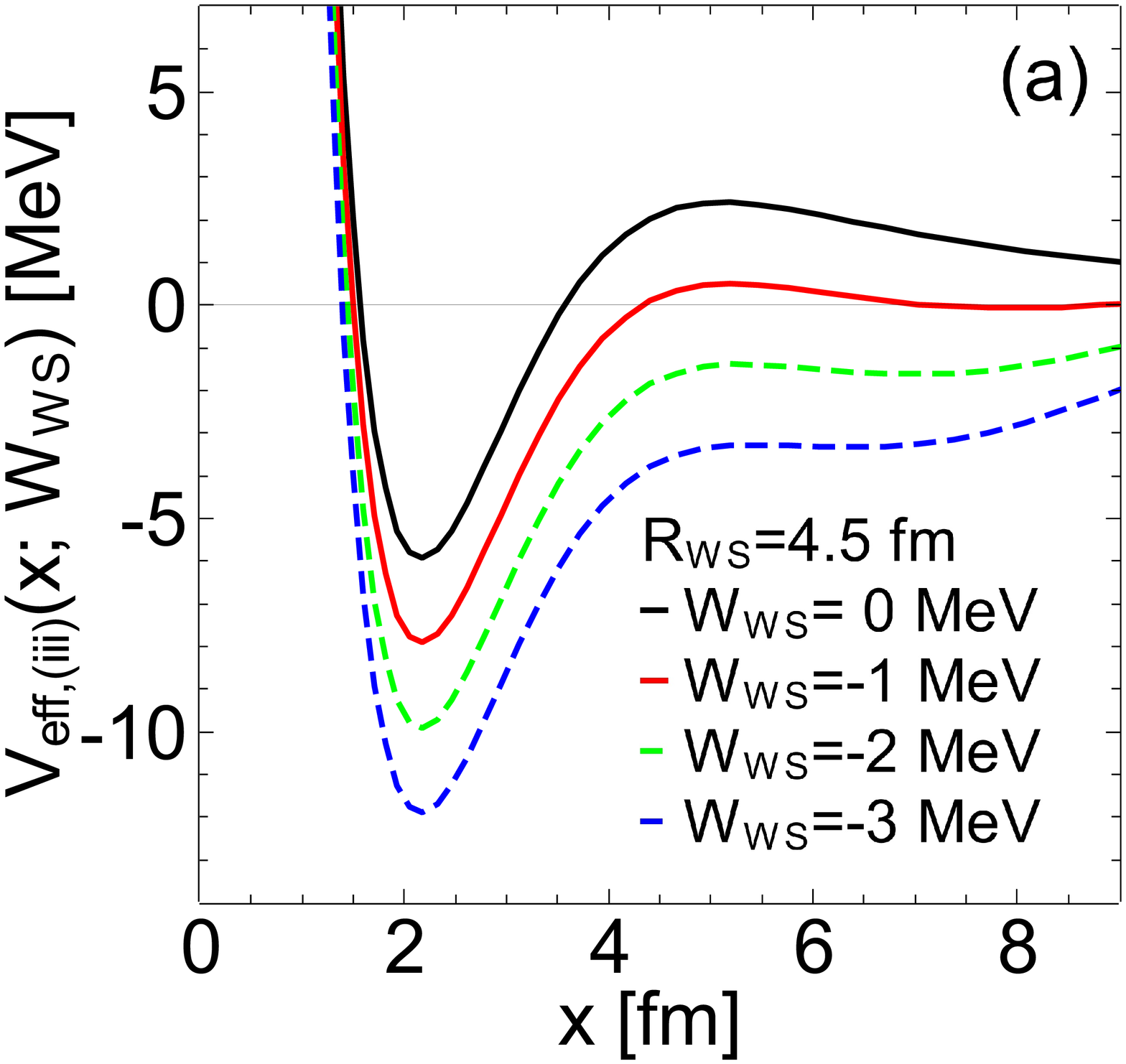}
\includegraphics[width=0.34\columnwidth,angle=-90]{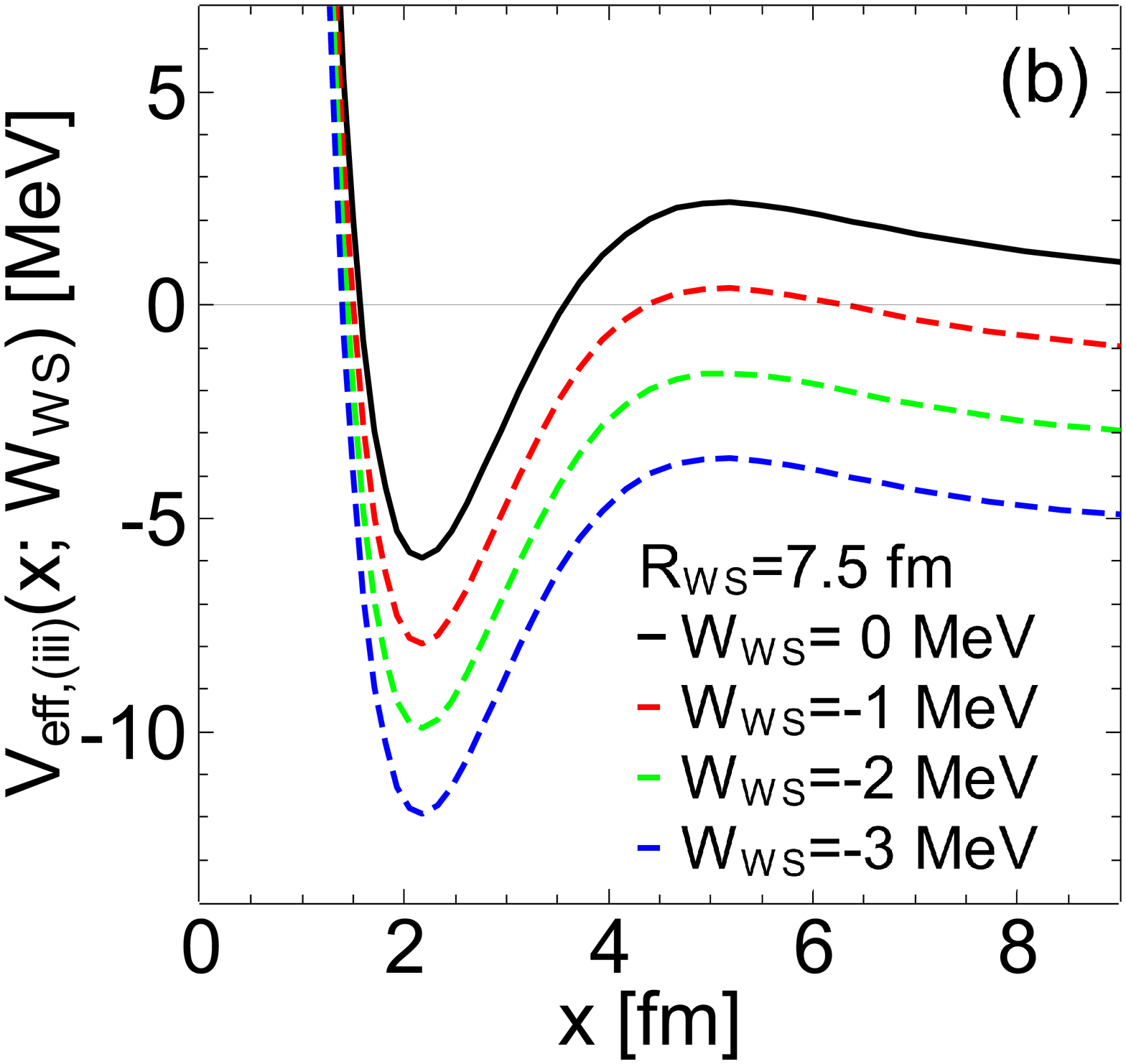}
\caption{(Color online) 
The effective potentials for (a) $R_{\mathrm{WS}}=4.5$ fm and  (b) $R_{\mathrm{WS}}=7.5$ fm. 
Black curves are for  $W_{\mathrm{WS}}=0$ MeV, 
red curves for  $W_{\mathrm{WS}}=-1$ MeV, 
green curves for  $W_{\mathrm{WS}}=-2$ MeV, and 
blue curves for  $W_{\mathrm{WS}}=-3$ MeV.
The meaning of solid and dashed curves is explained in the text.
\label{fig7:v-effective}
}
\end{figure}

The effective potentials $V_\mathrm{eff} (x;W_{\mathrm{WS}})$ for $W_{\mathrm{WS}}$ between $-3$ MeV and 0 MeV taking the range parameter of $R_{\mathrm{WS}}=4.5$ fm and $R_{\mathrm{WS}}=7.5$ fm are displayed in Fig. \ref{fig7:v-effective}.
In the figure, the solid curves indicate the potentials  for which no bound state exists, and dashed curves do those  for which a bound state exists. 

As the attractive effect becomes larger, the potential pocket spreads rapidly with vanishing barrier because the range of the trapping potential is longer than that of the Gaussian potential.
In other words, as the attractive effect is reduced, the barrier appears at positive energy, which may cause an extra repulsive effect that does not exist for the bound states.

\begin{figure}[tb]
\includegraphics[width=0.65\columnwidth,angle=-90]{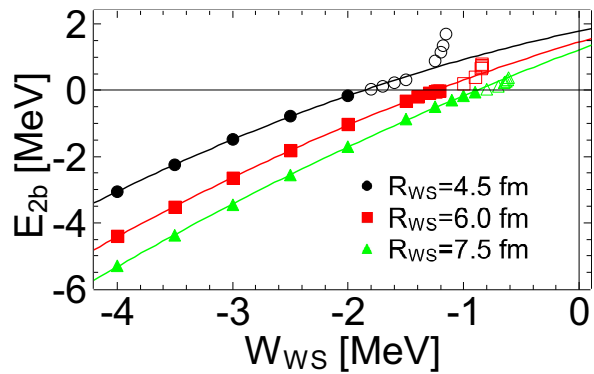}
\caption{(Color online) 
$W_{\mathrm{WS}}$ dependence of the energy of $2n$ p-wave state by the effective potential Eq. (\ref{eq:V-eff}) with  $v_\mathrm{G}=-50$ MeV.
Black circles, red squares, and green triangles denote calculations by the trapping potentials with $R_\mathrm{WS}=4.5$ fm, 6.0 fm, and 7.5 fm. 
Curves are obtained by fitting the binding energies.
\label{fig8:gauss50-P-WS}
}
\end{figure}

This extra repulsive effect is demonstrated in Fig. \ref{fig8:gauss50-P-WS}, which shows $W_{\mathrm{WS}}$ dependence of calculated values of the two-body energy for $R_{\mathrm{WS}}= 4.5$ fm, 6.0 fm, and 7.5 fm.
The dependence of the energy on $W_{\mathrm{WS}}$ at bound state region is described by 
a quadratic polynomial,
 and leads to a positive energy at  $W_{\mathrm{WS}}=0$ MeV.
However, soon after getting into the continuum region, the dependence is quite different from that at bound state region and 
the energy increases more than expected from the fitting, which indicates that the attractive effect becomes weak rapidly.
Because of the rise up at continuum region, the extrapolation  is no more reliable.

\section{Summary}
\label{sec:summary}

Continuum states of the $3n$ system are studied with the response function for the transition from ${}^3$H to $3n$ continuum state by an isospin excitation operator. 
We observe that the response function calculated with AV18 $nn$ potential does not reveal any resonance peak.

In view of the recent discrepancy in $3n$ calculations on the existence of $3n$ resonance state, I have examined three methods to bring an attractive effect to make the $3n$ system bind for extrapolating  the $3n$ energy: to enhance a component of the $nn$ potential, to introduce a three-body force, and to add an external attractive trapping potential. 
The first two methods are consistent with the non-existence of $3n$ resonance state. 
In the last case, the attractive effect unusually reduces the exterior barrier caused by the P-wave centrifugal potential, which makes the extrapolations using calculated $3n$ binding energies difficult.  
The reason for the unsuccessful extrapolation for the trapping method is due to the longer range trapping potential to destroy the potential barrier.  
This defect occurs in general, and the trapping method should be used carefully in studies of resonance states of few- and many-body systems.


\end{document}